\documentclass[
aip,
amsmath,
amssymb,
 reprint, 
author-year
]{revtex4-1}
\usepackage{CJKutf8}
\usepackage{graphicx}
\usepackage{dcolumn}
\usepackage{bm}
\usepackage{makecell}

\usepackage{graphicx}
\usepackage{dcolumn}
\usepackage{bm}
\usepackage{booktabs}
\usepackage[mathlines]{lineno}
\usepackage{appendix}

\usepackage{algorithm}%
\usepackage{algorithmicx}%
\usepackage{algpseudocode}%
\usepackage{listings}%

\usepackage{algorithm}
\usepackage{amsmath}
\usepackage{algpseudocode}
\usepackage{float}
\usepackage{listings}%
\usepackage{color}
 \usepackage{enumitem}
 \usepackage{graphicx}
\usepackage{subcaption}

\usepackage[utf8]{inputenc}
\usepackage[T1]{fontenc}
\usepackage{mathptmx}
\usepackage{etoolbox}
\modulolinenumbers[5]

\makeatletter
\def\@email#1#2{%
 \endgroup
 \patchcmd{\titleblock@produce}
  {\frontmatter@RRAPformat}
  {\frontmatter@RRAPformat{\produce@RRAP{*#1\href{mailto:#2}{#2}}}\frontmatter@RRAPformat}
  {}{}
}%
\makeatother
\begin{document}

\begin{CJK*}{UTF8}{gbsn}

\preprint{AIP/123-QED}

\title{DamFormer: Generalizing Morphologies in Dam Break Simulations Using Transformer Model}

\author{Zhaoyang Mu(牟昭阳) $^\dagger$}

\thanks{$^\dagger$ means the two authors contributed equally}
\affiliation{College of Artificial Intelligence, Dalian Maritime University, Dalian, China}
\author{Aoming Liang(梁敖铭) $^\dagger$}
\thanks{$^\dagger$ means the two authors contributed equally}
\email{liangaoming@westlake.edu.cn}
\affiliation{Zhejiang University-Westlake University Joint Training, Zhejiang University, Hangzhou, China}
\affiliation{ School of Engineering, Westlake University, Hangzhou, China}
\author{Mingming Ge (葛明明)} 
\altaffiliation[Also at ]{Institute of Advanced Technology, Westlake Institute for Advanced Study, Hangzhou, China}
\affiliation{Faculty of Science and Technology, Beijing Normal Univerisity-Hong Kong Baptist
University United International College, Zhuhai, China}
\affiliation{Key Laboratory of Coastal Environment and Resources of Zhejiang Province, School of Engineering, Westlake University, Hangzhou, China}
\author{Dashuai Chen (陈大帅)} 
\affiliation{Key Laboratory of Coastal Environment and Resources of Zhejiang Province, School of Engineering, Westlake University, Hangzhou, China}
\author{Dixia Fan (范迪夏)} 
\altaffiliation[Also at ]{Institute of Advanced Technology, Westlake Institute for Advanced Study, Hangzhou, China}
\affiliation{Research Center for Industries of the Future, Westlake University, Hangzhou, China}
\affiliation{Key Laboratory of Coastal Environment and Resources of Zhejiang Province, School of Engineering, Westlake University, Hangzhou, China}

\author{Minyi Xu(徐敏义)} \email{xuminyi@dlmu.edu.cn} 

\affiliation{State Key Laboratory of Maritime Technology and Safety, Marine Engineering College, Dalian Maritime University, Dalian, China}


\begin{abstract}

The interaction of waves with structural barriers such as dams breaking plays a critical role in flood defense and tsunami disasters. In this work, we explore the dynamic changes in wave surfaces impacting various structural shapes—circle, triangle, and square—using deep learning techniques. We introduce the "DamFormer" a novel transformer-based model designed to learn and simulate these complex interactions. The model was trained and tested on simulated data representing the three structural forms. Additionally, we conducted zero-shot experiments to evaluate the model's ability to generalize across different domains. This approach enhances our understanding of fluid dynamics in marine engineering and opens new avenues for advancing computational methods in the field. Our findings demonstrate the potential of deep learning models like the DamFormer to provide significant insights and predictive capabilities in ocean engineering and fluid mechanics. 

\end{abstract}

\maketitle

\end{CJK*}

\section{Introduction}
\label{sec:Introduction}

Dam break problems \citep{xiong2011dam,ferrari2010three,ozmen2011dam}, characterized by the sudden release of a fluid confined by a barrier, represent a fundamental topic in fluid dynamics with significant practical implications in the environment and hydraulic engineering. Despite extensive studies \citep{brufau2000two,biscarini2010cfd,jain2018brief,liu2024effects,vasanth2024comparative}, the complexity of predicting the physical behavior post-dam failure presents ongoing challenges, primarily due to the highly transient and three-dimensional fluid nature. For example, the failure of the South Fork Dam in Pennsylvania, USA, in 1889 led to over 2,200 fatalities \citep{eghbali2017improving}. In conventional numerical models, the water height distribution over time and space during dam break flows is commonly represented by the Saint-Venant equations, shallow water equations, or the Navier-Stokes equations \citep{wang2000finite}. Traditional models often rely on simplified assumptions that may not fully capture the dynamics of fluid structures at different scales and environmental interactions \citep{khoshkonesh2021numerical,zhou2024experimental}. However, due to the constraints imposed by the Courant-Friedrichs-Lewy (CFL) condition, solving the governing equations with traditional numerical methods like finite volume \citep{aliparast2009two} and finite difference \citep{ouyang2015maccormack}, typically demands a significant number of iterations, leading to high computational costs for these models.

Traditional machine learning methods have already been applied in this field, including neural networks \citep{juan2022review}, fuzzy inference systems \citep{ebtehaj2014performance, liang}, and support vector machines \citep{haghiabi2017prediction}. Notably, \citet{seyedashraf2018novel} propose a radial-basis-function (RBF) neural network to predict the water levels associated with dam break simulation. Although the aforementioned methods can predict fluid dynamics, it is crucial to achieve rapid model predictions, especially for transfer across different geometries.

More recently,  deep learning has offered a promising approach to computational fluid dynamics \citep{brunton2020machine,pawar2021physics,lee2024parametric}, particularly through the utilization of architectures like transformers \citep{alkin2024universal}, and operator-based learning \citep{azizzadenesheli2024neural}. These models are adept at handling sequential data, making them potentially suitable for temporal and spatial sequence prediction in fluid simulations. In the dam-breaking field,  \citet{kabir2020deep} introduces a CNN (convolution neural network) to predict the two-dimensional flood inundation. \citet{deng2024spatio} propose a Long Short-Term Memory with Attention (LSTM-AT) model to predict the water height for dam break. \citet{li2023data} adopt the echo-state neural network to learn a wave propagation behavior in the dam-break flood. The success of artificial intelligence has demonstrated that neural networks can facilitate long-term time series forecasting and generalization at boundaries, dynamics essential for learning the solutions to complex time-dependent partial differential equations. \citet{zhou2022deep} consider the boundary condition to train a U‐Net in the complex flow paths. 

While these models in dam break are capable of representing the evolution of fluid states, they have not been evaluated on larger-scale datasets. In this study, we focus on three-dimensional simulations and provide the datasets. Secondly, previous studies have not assessed the models' ability to generalize across different datasets in the Zero-Shot \citep{torrey2010transfer,sohail2024transfer}. We employ the concept of transfer learning, pre-training on one type of boundary condition and then proceeding with Zero-Shot and further training.

Our study introduces a novel application of the transformer-based
model named DamFormer to generalize the morphological variations observed during dam break events \citep{yang2024influence,zhang2024physical}. By leveraging the transformer's encoder to learn the dynamics in the long data sequences, our model predicts the evolution of fluid boundaries and interactions over time with improved accuracy and generalization capability. We aim to train the model on datasets with rectangular and circular shapes while ensuring that it generalizes well to triangular shapes. Specifically, we aim for the model to maintain strong performance even when applied to triangular conditions, despite being trained primarily on rectangular and circular data. 

To our knowledge, DamFormer could significantly aid in better understanding and predicting the catastrophic events of dam failures, thus contributing to the design of safer water retention structures and more effective emergency response strategies.

\section{Methodology}
\label{sec:Methodology}
Our methodology is illustrated in the following diagram. In this section, we detail introduce the dataset, model, and learning objective.

\subsection{Dataset and solver Description}

\begin{figure}[h!]
  \centering
  \includegraphics[width=0.88\linewidth]{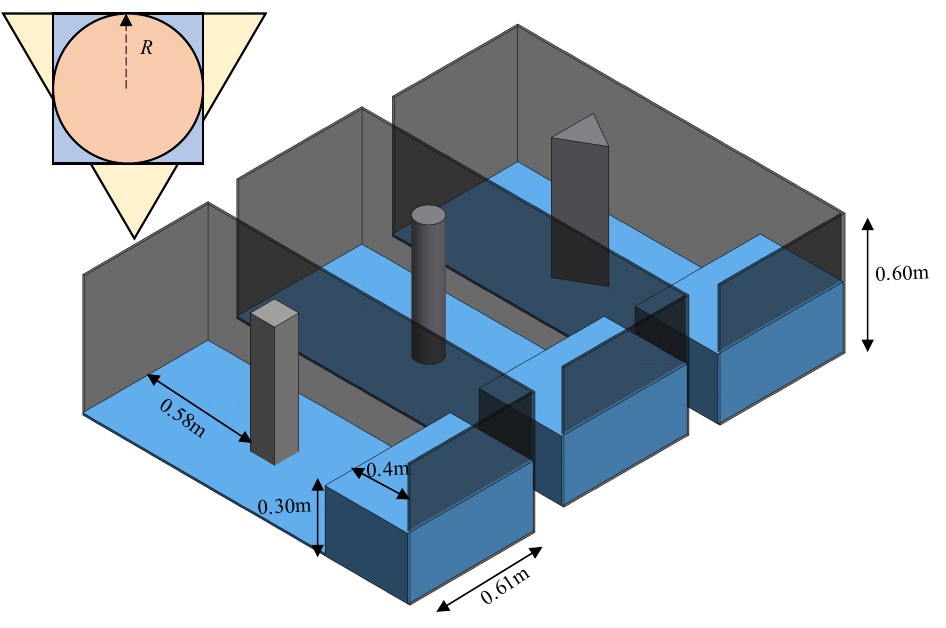}
  \caption{Physical boundary and initial condition are demonstrated through different boundary shapes.}
  \label{fig:physical}
\end{figure}

Wave impact is a critical issue in structural engineering. The shallow water equations are one of the simplest and most widely used sets of equations to simulate such problems. Initially, there is a body of water with a height of 0.3 meters behind the gate. At the start of the simulation, the gate is suddenly released, forming a wave that moves towards the structure. After the impact, the water flows forward until it reflects off a wall and impacts the opposite side of the structure again. 
Our model adopts the conservative form of the Shallow water equation:
\begin{align}
    \frac{\partial h}{\partial t} + \frac{\partial (hu)}{\partial x} + \frac{\partial (hv)}{\partial y} &= 0, \\
    \frac{\partial (hu)}{\partial t} + \frac{\partial (hu^2 + \frac{1}{2}gh^2)}{\partial x} + \frac{\partial (huv)}{\partial y} &= -gh\frac{\partial b}{\partial x}, \\
    \frac{\partial (hv)}{\partial t} + \frac{\partial (huv)}{\partial x} + \frac{\partial (hv^2 + \frac{1}{2}gh^2)}{\partial y} &= -gh\frac{\partial b}{\partial y}.
\end{align}
Here, \(h\) represents the water depth, \(u\) and \(v\) are the flow velocities in the \(x\) and \(y\) directions, respectively, \(g\) is the acceleration due to gravity, and \(b\) is the bed elevation.
The physical model employs a benchmark case from the COMSOL as shown in figure \ref{fig:physical}. For further details, please refer to the documentation. Our simulation data involved altering the size and shape of the geometry is shown in figure \ref{fig:physical}. We employ Monte Carlo sampling, conducting 40 simulations for each shape. The simulations span from 0 to 3 seconds, with outputs recorded at intervals of 0.1 seconds. The values of \( R \) are drawn from a truncated normal distribution with parameters \(\mu = 0.12\) and \(\sigma = 0.02\), constrained to the range \( 0.05 \leq R \leq 0.15 \). For the water depth, mesh grid points are obtained through post linear interpolation, resulting in a mesh of \(128 \times 128\) points in the spatial domain.

\begin{figure}[!htbp]
  \centering
  \includegraphics[width=0.95\linewidth]{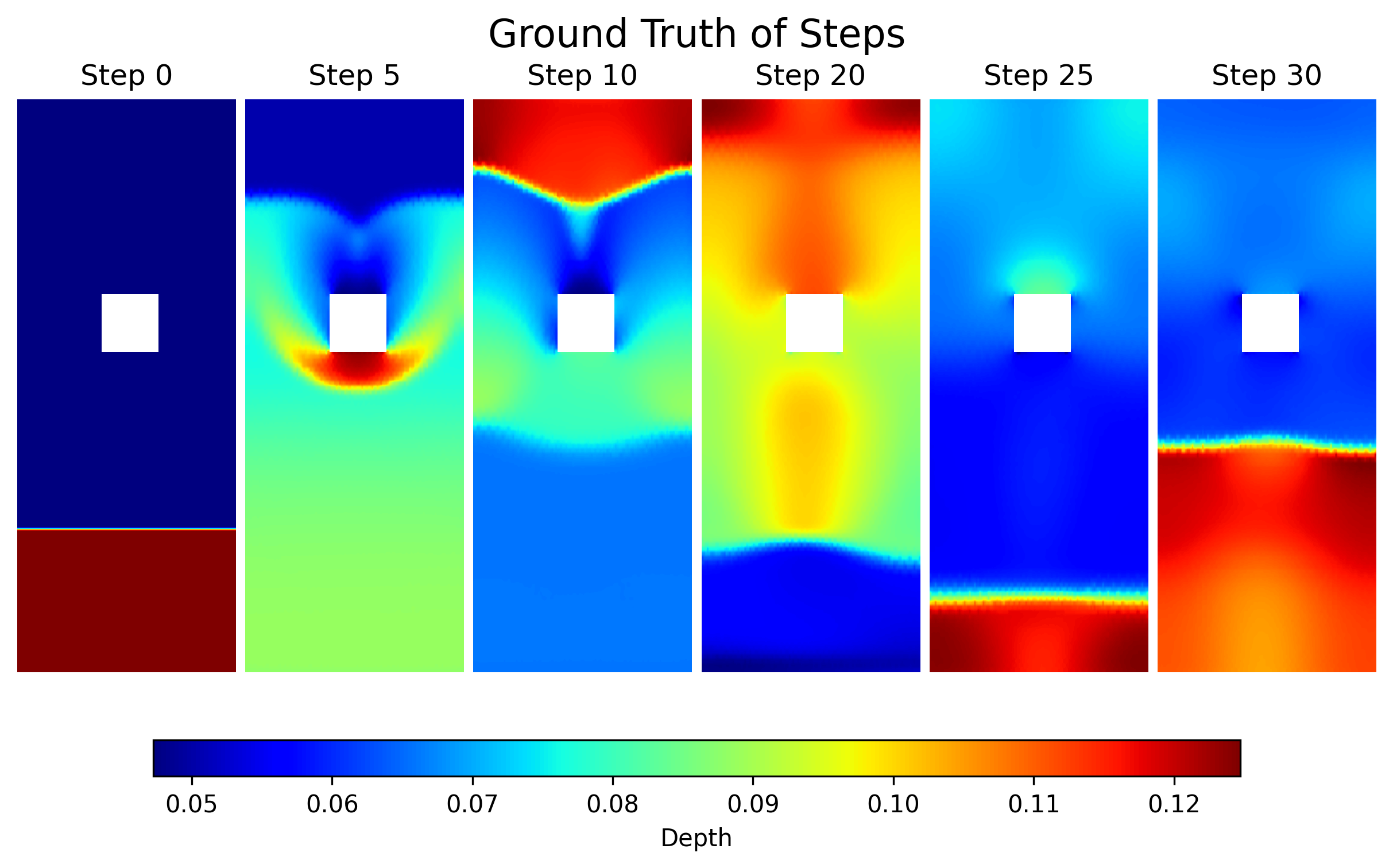}
  \caption{Numerical solution data on a circular boundary, from left to right, corresponding sequentially to the evolution at 0s, 0.5s, 1s, 2s, 2.5s, and 3s.}
  \label{fig:ground}
\end{figure}
\subsection{Model introduction}
Damformer processes data in a format similar to video. The input has the initial condition [batch, 1, $N_x$, $N_y$], while the output dimensions are [batch, $T-1$, $N_x$, $N_y$]. Our model processes initial conditions as input, which are then flattened and passed through a fully connected neural network layer. We describe the mathematical formulation of  DamFormer architecture:
\begin{equation}
\text{state}_0 \xrightarrow{f_\theta} \text{state}_1 \xrightarrow{f_\theta} \cdots \xrightarrow{f_\theta} \text{state}_{T}
\end{equation}
where the dimension of the depth state is the [batch, 1, $N_x$, $N_y$].  $f_\theta$ shows the evolution function. The primary task of the Damformer is to learn the evolution of states and to utilize the information about boundary conditions provided in the first step, enabling the model to learn the dynamic effectively.

\begin{figure*}[t]
  \centering
  \includegraphics[width=0.5\linewidth]{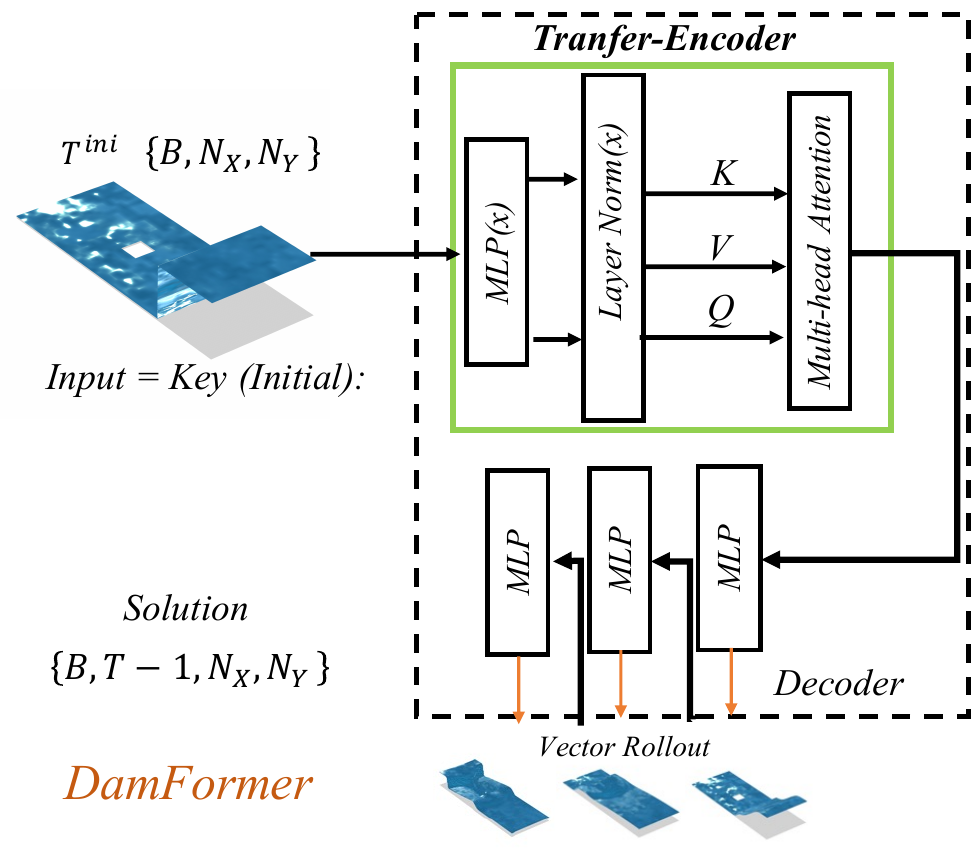}
  \caption{The architecture of DamFormer}
  \label{fig:AI}
\end{figure*}
Similar to the traditional transformer, Damformer has two parts: an encoder and a decoder. The encoder needs to employ dimensional transformations and attention mechanisms to compress the data effectively. The decoder is responsible for propagating the information processed by the encoder. The following illustrates the process of the Damformer.
\begin{itemize}
    \item \textbf{Input}: The input tensor \(X\) is first flattened into a vector of $N_x \times N_y$ dimensions for further processing.
    \item \textbf{Fully Connected Layer}: The flattened input passes through a fully connected layer:
    \begin{equation}
    X'' = W_f X' + b_f
    \end{equation}
    where \(W_f\) and \(b_f\) are the weights and bias of the layer.
    \item \textbf{Layer Normalization}: The output is then normalized using LayerNorm:
     \begin{equation}
    X''' = \text{LayerNorm}(X'')
    \end{equation}
    \item \textbf{Multi-Head Attention}: We compute multi-head self-attention by projecting the normalized output into query (\(Q\)), key (\(K\)), and value (\(V\)) vectors using learned projection matrices \(W^Q\), \(W^K\), and \(W^V\). Attention is computed as:
     \begin{equation}
    \text{Attention}(Q, K, V) = \text{softmax}\left(\frac{QK^T}{\sqrt{d_k}}\right)V
    \end{equation}
    where \(d_k\) is the dimensionality of the key vectors.
    \item \textbf{Encoder output}: The outputs from all heads are concatenated and projected back to the original space using a projection matrix \(W^O\):
    \begin{equation}
    \text{Output} = \text{Concat}(\text{head}_1, \text{head}_2, \ldots, \text{head}_h) W^O
     \end{equation}

     \item \textbf{Decoder Output} The decoder in the Damformer model is a fully connected neural network by iterative sequences based on the output of the encoder. 
     \begin{equation}
    x^0 = \text{Encoder Output}  
    \end{equation}
    \begin{equation}
    x^{t+1} = f(x^{t})  
    \end{equation}
    where $x^t$ represents the state of the sequence at time step \(t\), and \(f\) is the decoder function, projecting each subsequent state. 
    \begin{equation}
    \text{Outputs} = [x^{0}, x^{1}, \dots, x^{T}]  
    \end{equation}
\end{itemize}

This Output Sequence collects the states generated at each time step, forming the complete sequence that the model predicts.

\subsection{Learning Objective}
In the training stage, we use the relative L2 norm as the loss function. The objective is defined as follows. 

\begin{equation}
\text{Loss} = \frac{\left\| \text{Outputs}_{1:\text{end}} - \text{True}_{1:\text{end}} \right\|_2}{\left\| \text{True}_{1:\text{end}} \right\|_2}
\end{equation}
where True is the solution by the solver. 
  
\section{Results and Discussions}\label{sec:Discussion}
For our evaluation metrics, we chose the predicted RMSE values, along with the number of parameters and inference speed of different transformer models. The experiments were conducted with a transformer configured with the following hyperparameters.

\begin{table}[!htbp]
\centering
\caption{Hyper parameters of training stage, the DamFormer is based on Pytorch }
\label{tab:hypers}
\begin{tabular}{lc}
\toprule
Setting & Value  \\
\midrule
Batch size & 20   \\
learning rate  & $10^{-3}$ \\
Head of DamFormer & $4$ \\
Feed Forward & $256$ \\
Encoder Layer & $2$\\
Decoder Dimension & $256$\\
Optimizer  & $Adam$\\
Parameters($10^7$) & $9.9$ \\

\bottomrule
\end{tabular}
\end{table}
To distinguish the shapes of the obstacles in the model, we used them as a mask for differentiability and automatic differentiation. During the training process, we marked the values in these regions as a constant zero.

We observe that the adaptations from transfer learning vary with different boundary shapes of the model input. The transfer learning process will showcase interoperable attention maps.
\subsection{Performance in the circle boundary}

We selected 40 Monte Carlo simulations with circular boundaries for our experiments, using a training-to-testing ratio of 7:3.  The ground truth and the predictions from the DamFormer are in Figures 1 and 2.

\begin{figure}[!h]
  \centering
  \includegraphics[width=1\linewidth]{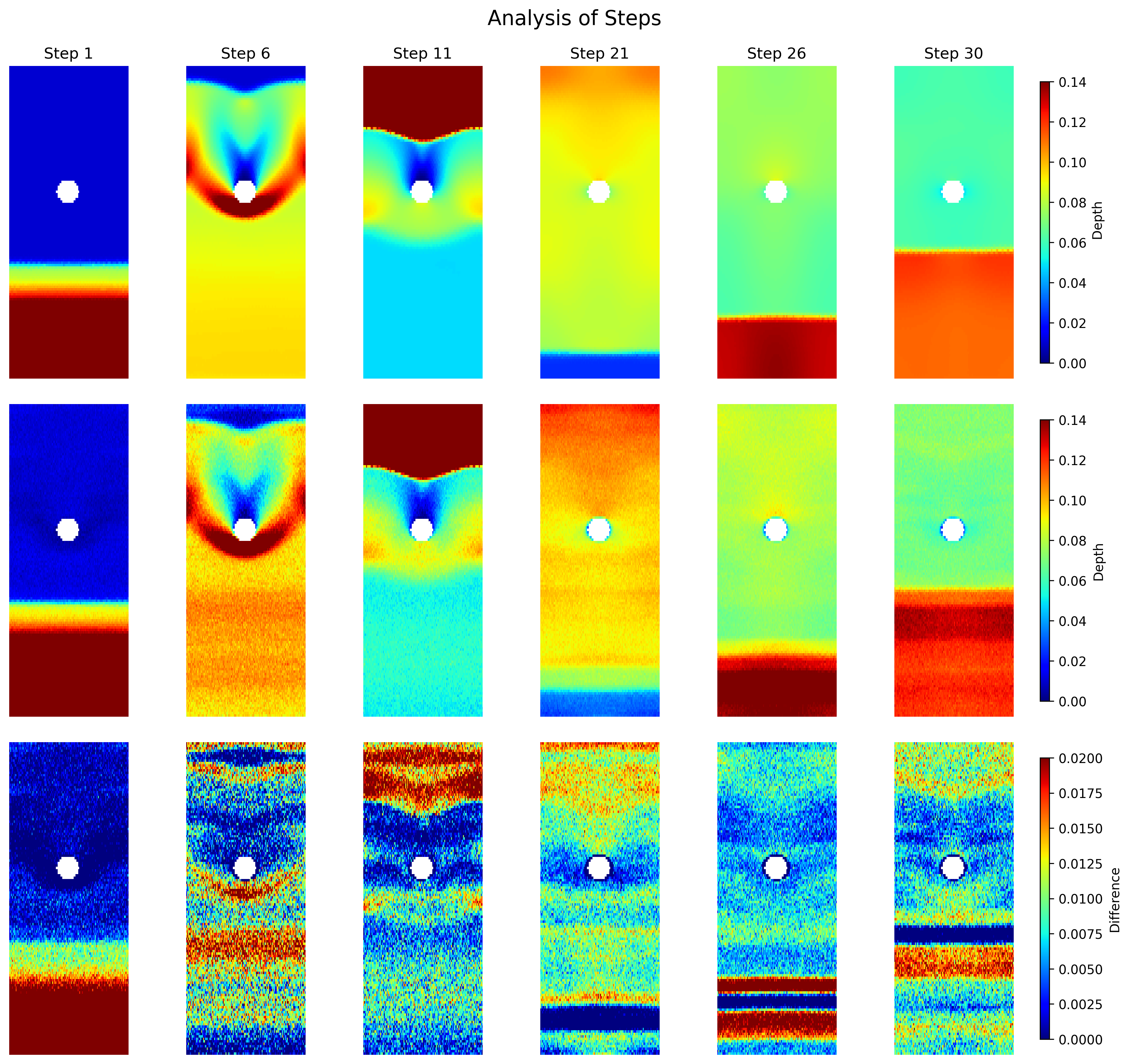 }
  \caption{Comparison between the ground truth and output of Damformer}
  \label{fig:AI}
\end{figure}

\subsection{Performance in the square boundary}

The results for the rectangular boundary area are shown in the following figure. As can be seen, the predictive capability of the Damformer is quite as good as the circle boundary.
\begin{figure}[!h]
  \centering
  \includegraphics[width=1\linewidth]{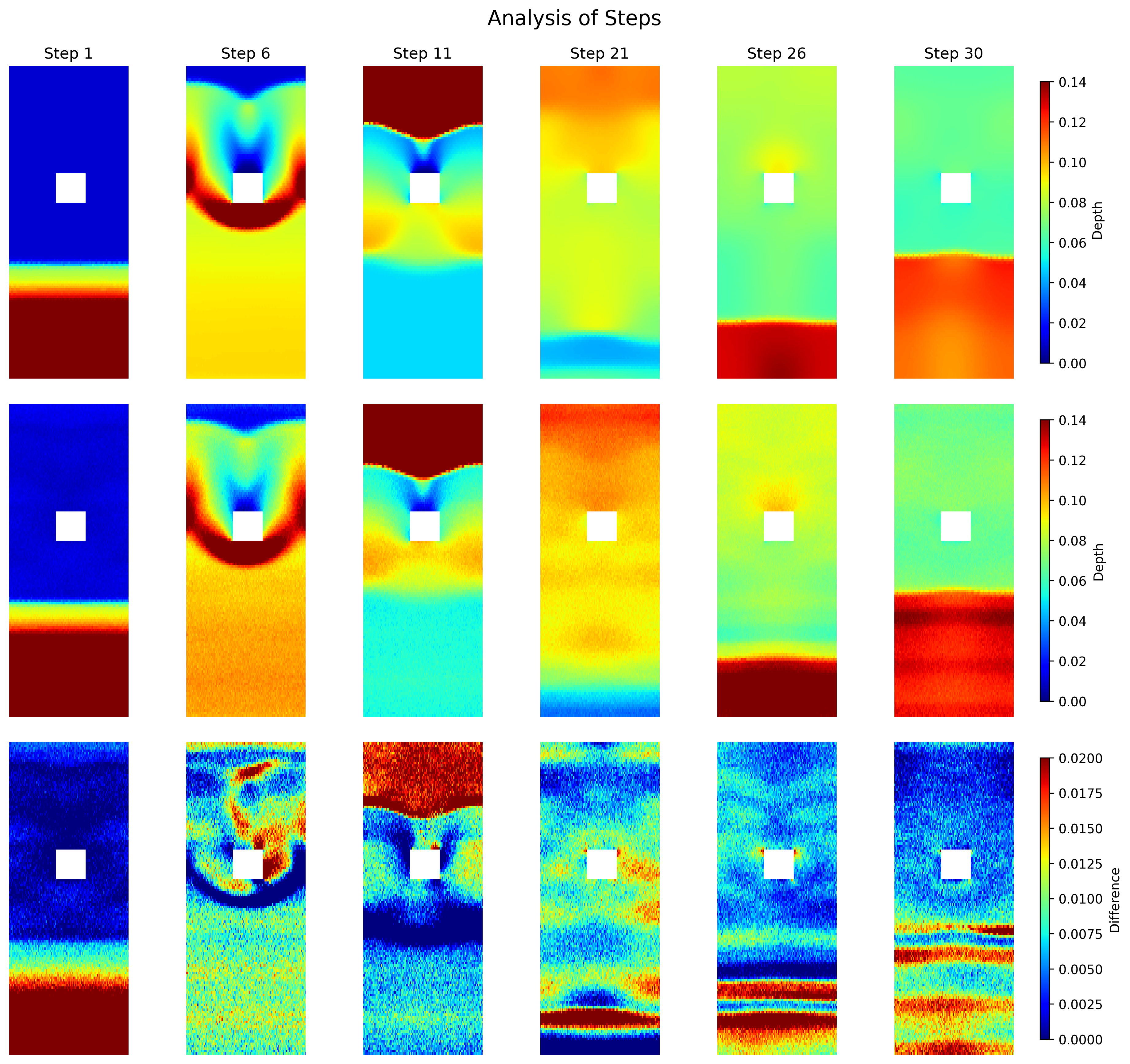 }
  \caption{Comparison between the ground truth and output of Damformer}
  \label{fig:AI}
\end{figure}

\subsection{Performance in the triangle boundary}

As seen in the figure below, the triangular boundary may perform slightly worse than the two results above. The predictions behind the obstacle are not ideal, which may be due to issues with the mask data. However, the trend of the data is consistent with the process of fluid evolution.
\begin{figure}[h!]
  \centering
  \includegraphics[width=0.95\linewidth]{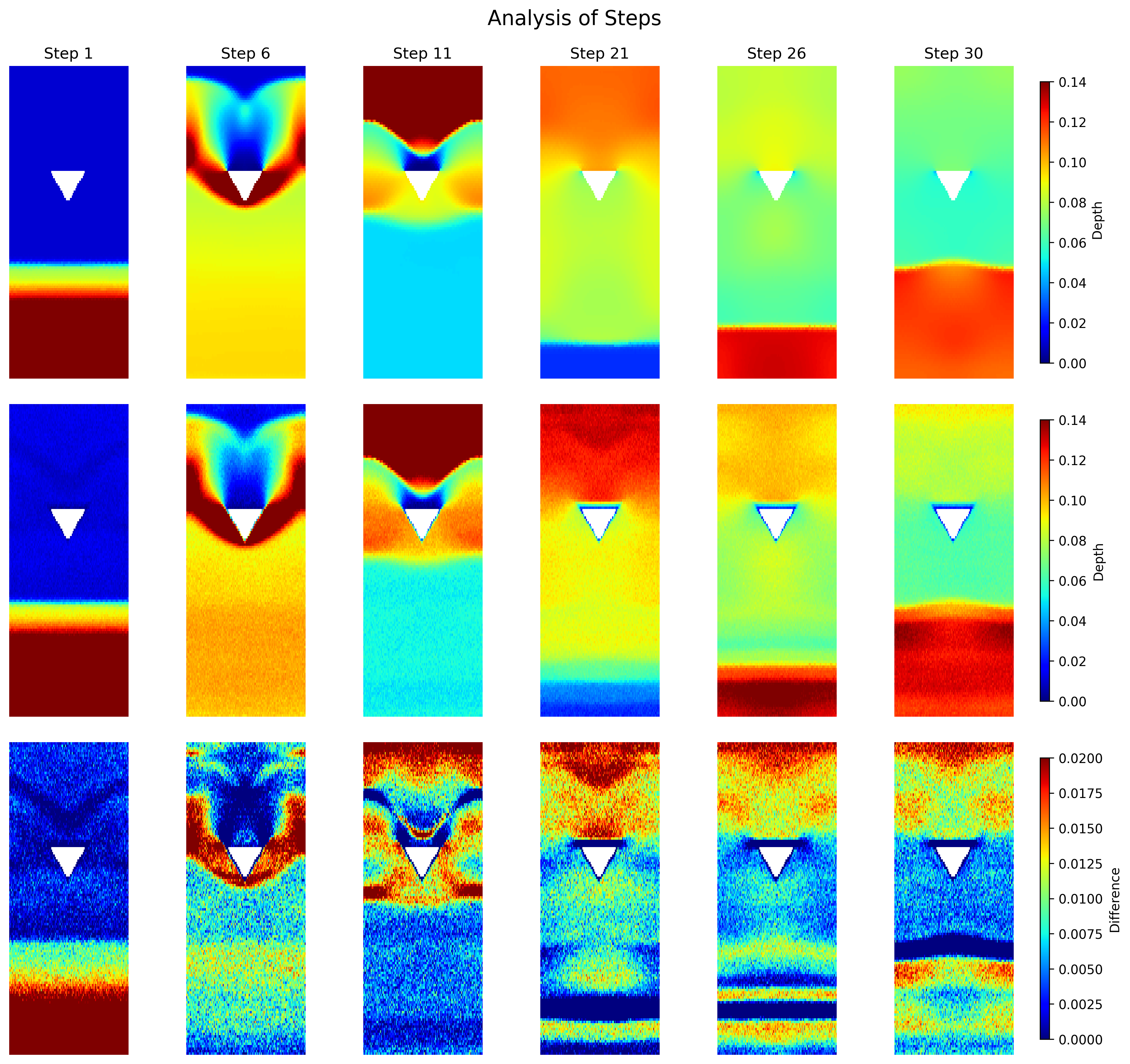}
  \caption{Comparison between the ground truth and output of Damformer}
  \label{fig:AI}
\end{figure}

\subsection{Performance on transfer learning results on generalized boundary conditions}

In this section, we primarily aim to utilize transfer learning to adapt a model trained on one boundary condition with an encoder, assuming that the encoder is sufficiently trained, to another boundary condition. We define the transfer error as the relative L2 norm on the test set like Zero-Shot. Subsequently, we train the model specific to the new boundary conditions and quantify the error named final Error.

\begin{table}[!htbp]
\centering
\caption{Results on statistical tests indicated significant differences in the transfer learning performance among the different shapes: triangle (T), square (S), and circle (C). }
\label{tab:hypers}
\begin{tabular}{lcc}
\toprule
Setting & Final error & Transfer error  \\
\midrule
T2C & 0.082 & 0.24\\
T2S &0.085 & 0.26\\
C2T & 0.093 & 0.11\\
C2S & 0.080 & 0.13\\
S2C & 0.081 & 0.10\\
S2T & 0.092 & 0.21\\
T&0.090&None\\
C&0.072 &None \\
S& 0.083&None\\

\bottomrule
\end{tabular}
\end{table}
\begin{figure}[h!]
  \centering
  \includegraphics[width=0.95\linewidth]{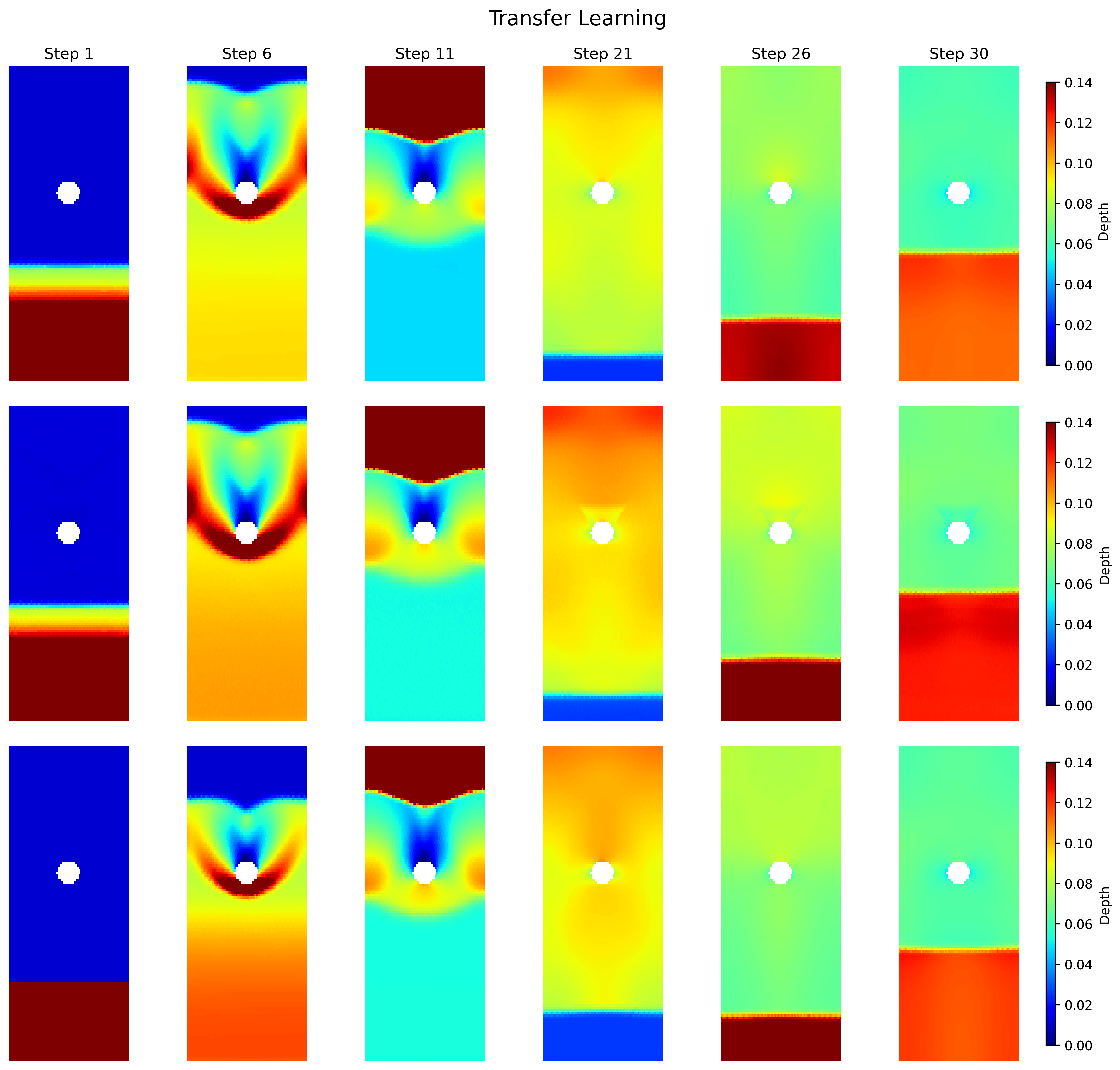}
  \caption{Comparison on T2C at zero-shot test, first row displays the ground truth, the second row shows the results of zero-shot learning, and the third row presents the outcomes after further training}
  \label{t2c}
\end{figure}

From the table above, it is evident that the model's zero-shot transfer from circles to squares performs the best. In contrast, the performance of transferring from triangles to circles is relatively poor, indicating that additional training is required to reduce errors. This suggests that the encoder may not handle the boundaries of triangles effectively. By qualitatively examining the attention maps, we can observe the differences between them.
\section{Concluding remarks}
\label{sec:Conclusions}

In this study, we leveraged the transformer architecture \citep{vaswani2017attention} to perform roll-outs for traditional numerical solutions of dam break simulations. We conducted tests across various boundary conditions and developed a transfer learning framework capable of adapting to different boundaries. Our experimental results revealed that our model achieves a fivefold speed improvement over traditional numerical solutions. Moving forward, we will investigate the effects of structures in large-scale oceanic environments.
Additionally, to further evaluate the generalizability of our transfer learning approach, we will test more physical models.
\section*{Acknowledgements}
The project is supported by the Priority Postdoctoral Projects in Zhejiang Province (ZJ2023023) and
GuangDong Basic and Applied Basic Research Foundation (2024A1515010711). The first author would like to express gratitude to the Westlake University Education Foundation for their generous support.
\section*{Data Availability}
The datasets and models during the current study are available at https://github.com/liangaomng/pof-for-dam.git.

\setcitestyle{authoryear,open={((},close={))}} 

\bibliography{reference}

\end{document}